\documentclass[aps,prd]{revtex4}
\usepackage{amssymb,amsfonts,latexsym}

\usepackage{graphicx}

\def\be{\begin{eqnarray}}
\def\en{\end{eqnarray}}
\def\bea{\begin{eqnarray}}
\def\ena{\end{eqnarray}}

\draft

\begin{document}

\title{Signal based vetoes for  the detection of gravitational waves from inspiralling compact binaries.}

\author{S. Babak}
\affiliation{School of Physics and Astronomy, Cardiff University,
Cardiff CF24 3YB, UK}
\author{H.~Grote, M.~Hewitson, H.~L{\"u}ck}
\affiliation{Max-Planck-Institute, AEI, Callinstra{\ss}e 38, 30167-Hannover, Germany}
\author{K.A.~Strain}
\affiliation{Department of Physics and Astronomy, University of Glasgow,
Glasgow, G12 8QQ}
\begin{abstract}
The matched filtering technique is used to search for gravitational wave signals 
of a known form
in the data taken by ground-based detectors. However, the analyzed data contains a number of artifacts arising from various broad-band transients (glitches) of instrumental or environmental origin which can appear with high signal-to-noise ratio on the matched filtering 
output. This paper describes several techniques to discriminate genuine events from the false ones, based on our knowledge of the signals we look for. Starting with the $\chi^2$ discriminator, we show how it may be optimized for free parameters. We then introduce several alternative vetoing statistics and discuss their performance using data from the GEO\,600 detector. 
\end{abstract}

\maketitle

\section{Introduction}

The first generation of gravitational wave detectors is either already 
online and gathering scientific data (LIGO \cite{LIGO}, GEO\,600 \cite{GEO}, 
TAMA \cite{TAMARev}) or about to start taking data (VIRGO \cite{VIRGO}). 
LIGO and GEO\,600 have successfully completed 
several short data taking runs (so called science runs) in coincidence \cite{S1_1, S1_2}. TAMA has accumulated over 2000 hours of data  \cite{TAMA, TAMAchi}
 and quite a big portion of this data was taken in coincidence with LIGO and GEO\,600. All detectors are currently in the commissioning stage and are steadily approaching
 their design sensitivities. Improvements in the performance of the detectors are carried out in several directions:
(i) sensitivity improvements (tracing and reducing noise level from different subsystems) (ii) increasing duty cycle (time spent in acquiring the data suitable for astrophysical analysis as a fraction of the total operational time), and (iii) improving the data quality (stationarity).

However, at the present state the data is neither stationary nor Gaussian over 
time scales greater than few minutes. The detector output contains various spurious transient events. Unfortunately, the output of an optimal filter reflects these events, especially various glitches.
By glitch here we mean a short duration spurious transient (of almost delta-function shape) with a broad band spectrum that leads to a high signal-to-noise ratio (SNR) at the output of matched filtering. Distinguishing  these events from the real events of astrophysical origin and dropping them out of consideration is called vetoing. 
In addition to the main gravitational wave channel, interferometers record a large volume of auxiliary data from environmental
monitors and various signals from the many detector subsystems.
These monitors help to find correlations between abnormalities in environmental or in instrumental
behaviour and events in the strain channel with high SNR.  The transients which correlate both in the strain and auxiliary channels (occure in both within a coincidence window) can be discarded on the ground of noise coupling between the strain channel and detector's subsystems (provided we understand the physical reasons for such a coupling mechanism). This is what is regarded as instrumental vetoes. 
The instrumental vetoes are helpful for removing some fake events, however, it is
not enough. We have other events which are of artificial nature, but the
information which would help us to remove these events either was not recorded or is not recognised. So in addition to instrumental vetoes, we need to apply signal based vetoes:
 vetoes which are based on our knowledge about a signal's shape in the frequency- and/or time-domain. For signal based vetoes, we need to construct a statistic which helps us to 
discriminate false signals from the true ones. The $\chi^2$ time-frequency discriminator suggested in \cite{Ballen} is an example of such a statistic. This vetoing statistic is used in a search for gravitational waves from the binary systems consisting of two compact objects (Neutron Stars (NS), Black Holes (BH),...) orbiting around each other in an inspiralling trajectory due to loss of orbital energy and angular momentum through gravitational radiation. A lot of effort has been put into modeling the waveform from coalescing binaries \cite{PPN, EOB, EOBspin, DIS}. 
The waveforms (often referred to as chirps) are modelled with reasonable accuracy, so that matched filtering can be employed to search the data for these signals. In the case of the $\chi^2$ discriminator, we
use the time-frequency properties of the chirp in order to discard (to veto out) any 
spurious event which produces an SNR above a preset threshold on the matched filter output. The performance of $\chi^2$ might depend on the number of bins used in computing the statistic. 

In this paper we suggest a possible way to optimize the $\chi^2$ discriminator for the number of bins. We use software injections (adding simulated signals) into data taken by the GEO\,600
detector during the first science run (S1) in order to study the distribution of the $\chi^2$ 
statistic for simulated signals and for noise-generated events. The optimal number of bins
is the one which maximizes detection probability for a given false alarm rate. This method
is quite generic and can be used for tuning any vetoing statistic which depends on one or several parameters.

Though the $\chi^2$ discriminator works reasonably well, it is still desirable to have additional independent signal based vetoes, which would either increase our confidence or improve our ability to separate genuine events from spurious ones. 
Some investigations have been already made in this direction \cite{TAMAchi, Baggio}.  In addition to signal based vetoes, a heuristic veto method was suggested in \cite{Shawhan}. It is based on counting the number of SNR threshold crossings within a short time window.

In this paper we suggest several new signal based statistics which can compliment $\chi^2$ or enhance its performance. We introduce a statistic inspired by the Kolmogorov-Smirnov 
``goodness-of-fit'' test \cite{Kolmogorov}, we call it the $d$-statistic. 
We derive its probability distribution
function in the case of signals buried in Gaussian noise. We have also suggested a few other $\chi^2$-like and $d$-like statistics and show that their combination could increase vetoing efficiency even further. 
 
Throughout the paper we have used the following assumptions and simplifications.
We shall assume that the waveforms used in our simulations, ``Taylor'' 
approximants (t1) at second Post-Newtonian order in the notations used in \cite{DIS}, are the exact representation of the astrophysical signal. 
The study performed in this paper is not restricted by the waveform model and could be repeated for any other model at the desirable Post-Newtonian order. 
The waveforms depend on several parameters, some of these parameters are intrinsic to the system like the masses and spins, while others are extrinsic like the time and phase of arrival of the gravitational wave signal. To search for such signals we use a bank of templates, which can be seen as a grid in the parameter space \cite{Owen}. Separation of templates in the parameter space is defined by the allowed loss in the SNR (or equivalently by a loss in the detection probability). The detector output is usually filtered through a bank of templates for parameter estimation \cite{Cutler, Bala}. For the sake of simplicity we have used a single template with parameters identical, or very close, to those of the signal used in the Monte-Carlo simulation described in Section~\ref{III}.

This paper is structured as follows.
We start in Section~\ref{II} by recalling the widely used \cite{40m, TAMAchi} $\chi^2$ time-frequency discriminator \cite{Ballen}. 
In Section~\ref{III}, we describe the method to optimize the $\chi^2$ veto for the number of bins. Though we show its performance for $\chi^2$ optimization, the method is applicable to any discriminator which depends on some free parameters.
Section~\ref{IV} is dedicated to alternative vetoing statistics. 
There we start with the $d$-statistic, then we show few more examples of $d$- and $\chi^2$-like statistics ($\hat{d}$ and $\hat{r}^2$ correspondingly) which can potentially increase the vetoing efficiency further.
For instance we show that the combination of $\hat{d}$ and $\hat{r}^2$
statistics (namely their product) give the best performance for a day's worth GEO\,600 data. We summarize main results in the concluding Section~\ref{V} and some detailed derivations are given in Appendix~\ref{App1}.

\section{Conventions and $\chi^2$ discriminator}\label{II}

In this Section we introduce the notation which will be used throughout 
the paper and we reformulate the $\chi^2$ discriminator \cite{Ballen} using
 new notations. This should be useful in the following sections where 
we discuss $\chi^2$ optimization and alternative signal-based vetoing
statistics.

Throughout this paper we assume that the signal is of a known phase with known time of arrival without loss of generality. Indeed, we can use phase and time of arrival taken from the maximization of SNR. Alternatively, one can extend the derivations below in a manner similar to \cite{Ballen} to deal with the unknown phase. 

The detector output sampled at $t_j = j\Delta t $ is denoted by $x(t_j) = n(t_j) + As(t_j)$, where $n(t_j)$ is  noise and $s(t_j)$ is a signal, which corresponds to the gravitational wave of amplitude $A$. Since we will be working mainly in the frequency domain,
we use tilde-notation for a Fourier image of the time series: $\tilde{x}(f_k) =  \tilde{n}(f_k) + A\tilde{s}(f_k)$. 
The discrete Fourier transform is defined as
$$
\tilde{x}(f_k) = \sum_{j=0}^{M}x(t_j) e^{-2\pi i t_j f_k},
$$
where $f_k = {k\over{M\Delta t}}$, and $M$ is a number of points.

In order to introduce the $\chi^2$ discriminator we need to define the following quantities

\bea
S_i &=& 2\sum_{f_k = F_{i-1}}^{F_i} \frac{\tilde{s}(f_k)\tilde{s}^*(f_k)}{S_n(f_k)}\Delta f,\;\; 
i= 1,\ldots, N;\;\;\;\;
S = 2\sum_{f_k = F_{0}}^{F_N} \frac{\tilde{s}(f_k)\tilde{s}^*(f_k)}{S_n(f_k)}\Delta f,\\
Q_i &=& \sum_{f_k = F_{i-1}}^{F_i} \;\left(\frac{\tilde{x}(f_k)\tilde{s}^*(f_k)}{S_n(f_k)}   +  
c.c.\right)\Delta f,\;\;
i= 1, \ldots, N;\;\;\;\;
Q = \sum_{f_k = F_{0}}^{F_N} \left(\frac{\tilde{x}(f_k)\tilde{s}^*(f_k)}{S_n(f_k)}  + 
c.c.\right)\Delta f.
\ena
Note that this notation is different from that used in \cite{Ballen}. Here the one-sided noise 
power spectral density (PSD), $S_n(f_k)$, defined as
$$ \frac{M}{2\Delta t}S_n(|f_k|) \delta_{kk'} = E(\tilde{n}(f_k)
\tilde{n}^*(f_{k'})), 
$$
is assumed to be known, $c.c.$ as well as  ``*'' mean complex conjugate
and $\Delta f = {1\over{M\Delta t}}$.
We have chosen to work with discrete time and frequency series to be close to reality. Here and after we use $E(...)$ for the average over ensemble and $var(...)$ for the second moment of the distribution.
The frequency boundaries $F_0, F_N$ 
correspond to the frequency at which the gravitational wave signal 
enters the sensitivity band of the instrument \footnote{ It is so 
called frequency of the gravitational wave signal at the ``time of arrival''.} and the frequency at the last stable orbit, $F_N = f_{lso}$ \footnote{ If $f_{lso}$ is beyond the Nyquist frequency,
$f_{Ny} = {1\over{2\Delta t}}$, $F_N$ should be taken as $f_{Ny}$.} (sometimes it is also referred to as the frequency at the innermost stable circular orbit). 
In this notations $Q$ corresponds to the SNR (up to a numerical factor which does not play any role in the further analysis) and $Q_{i}$ is a part of the total
SNR accumulated in the frequency band between $F_{i-1}$ and $F_{i}$.
We choose a normalization for the templates so that $S=1$. Let us emphasize again,
that we have assumed that we know the phase and time of arrival, so they are
incorporated in the definition of the waveform $s(t_i)$.

For the $\chi^2$ discriminator, we choose the frequency bands ({\it bins})
$F_1,\ldots, F_N$, so that there is an equal power of signal in each band:
$S_i = S/N = 1/N$. Then the $\chi^2$ discriminator can be written in the notations adopted here as follows

\bea
\chi^2 = N\sum_{k=1}^{N} (Q_i - Q/N)^2.
\label{chi2}
\ena
If the detector noise is Gaussian, then the above statistic obeys a $\chi^2$ distribution with $N-1$ degrees of freedom.  The main idea behind the $\chi^2$ discriminator is to
split the template $\tilde{s}(f)$ into sub-templates defined in different frequency bands, so that if the data contains the genuine gravitational wave signal, the contributions ($Q_i$) from each sub-template to the total SNR ($Q$) are equal ($E(Q_i) = A/N$). 

In the presence of a chirp in the data or if the data
is pure Gaussian noise, the value of $\chi^2$ is low 
$E(\chi^2) = N-1$.
However, if the data contains a glitch which is not consistent with the inspiral signal, then the value of $\chi^2$ is large.
This statistic is very efficient in vetoing all spurious events that cause large SNR in the matched filter output. It was used in the search pipeline for setting an upper limit on the rate of coalescing NS binaries \cite{S1_2}.

If we want to apply this vetoing statistic in a binary BH search we should do some modifications of the $\chi^2$ discriminator Eq.~(\ref{chi2}) to increase its efficiency. 
In practice, it might be difficult to split $S$ in bands of exactly the same power $S_i$ for signals from high mass systems, in other words it might be difficult to achieve $S_i = 1/N$ exactly. 
Indeed, the bandwidth of the signal from binary BH decreases with increasing total mass, $f_{lso} = 1/(6^{3/2}\pi m)$, where we used $G=c=1$, and $m=m_1+m_2$ is the total mass \footnote{We do not include merger and quasi-normal modes at the end of inspiralling evolution.}. 
In addition we work with a finite frequency resolution, which we might want to decrease to save computational time. Finally, the  accuracy of splitting the total frequency band depends on the number of bins.

Based on this we suggest a modification of the $\chi^2$ 
discriminator, which does not change it statistical properties,
but enhances its performance \cite{Ballen}. We introduce $p_i = S_i/S$ which is close to $1/N$, but not exactly equal to it. Then we should redefine $\chi^2$ statistic according to

\bea
\chi^2 = \sum_{i=1}^{N}\frac{(Q_i - p_iQ)^2}{p_i}
\label{chi2mod}
\ena
We refer to \cite{Ballen} for more details on this modification and
its properties.

\section{Optimization of vetoing statistic}\label{III}

 In this section we would like to present a method for optimizing
 parameter-based vetoing statistics. This 
method also helps to tune the veto threshold for a signal based 
statistic. Though the main focus in this section 
will be on the optimization of the $\chi^2$ statistic with respect to the number of bins, this method 
can also be applied to a general case (see Section~\ref{IV}).
 
First we need to define playground data. Playground data is a small subset of the available data chosen to represent the statistical properties of the whole data 
set \cite{Playground}. The main idea is to use software injections of the
chirps (adding simulated signals) into playground data and compare the distribution of $\chi^2$ for the injected signals and spurious events.
There is a trade off between the number of software injections: on the one hand we should not populate the data stream with
too many chirps as it will corrupt the estimation of PSD, on the other hand the number of injections should not be too small, so that we can accumulate sufficiently large number
of samples (``sufficiently large'' should be quantified, 
see \cite{Kay}). Another issue is the amplitude of injected signals: 
the amplitude should be realistic, which means
close to the SNR threshold used for the search. Parameters of the injected chirps (such as masses, spins, etc.) should be either fixed (optimization with respect to the particular signal) or correspond to the range of parameters used for templates in the bank.
 A generalization could be optimization with respect to several 
 (group of) signals and the use of different number of bins 
for different (set of) parameters. That could happen in reality: the
search for binary NS and binary BH might have different optimal number of bins.

\begin{figure}[tbh]
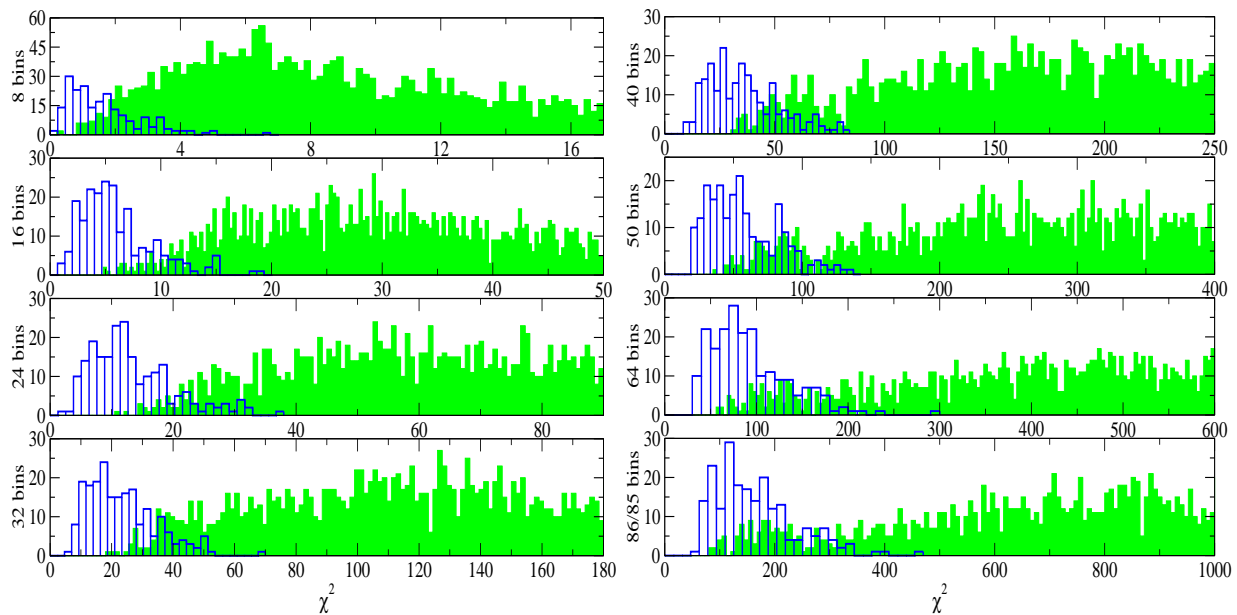

\resizebox*{.45\textwidth}{.45\textwidth}{\includegraphics{1_stat_hs.eps}}
\resizebox*{.45\textwidth}{.45\textwidth}{\includegraphics{2_stat_hs.eps}}
\caption{Distribution of $\chi^2$ for simulated signals (the histogram drawn by the solid line), and for spurious events in GEO\,600
S1 data (the shaded histogram). }
\label{chi21}
\end{figure}

To ease our way through we give an example of the optimization of $\chi^2$ for signals from the 
$5-5 M_{\odot}$ system. We injected a waveform with mass parameters $5.0-5.0 M_{\odot}$ and SNR=13 in each 
5th segment of analyzed data. Each segment was 16 seconds long. Then 2.5 hours of GEO\,600 S1 data was 
filtered through the template TaylorT1 (at 2-nd Post-Newtonian order) with mass parameters $5.04-5.04 M\odot$. The template TaylorT1 corresponds to ``t1'' in \cite{DIS}. By having a slight mismatch in masses of the system, we have tried to mimic a possible mismatch due to the coarseness of the template bank.
We have separated triggers which correspond to the injected signals 
from the spurious events by using a 5 msec window around the time of
injection. SNR threshold was chosen to be 6. Then we have produced 
histograms for $\chi^2$ distribution for injected/detected signals and for spurious events. This procedure was performed for different number of bins for $\chi^2$ statistic. One can see the results in Figure \ref{chi21}. The solid line histogram shows the distribution of $\chi^2$ for signals and the shaded histogram corresponds to the distribution of $\chi^2$ for spurious events with $SNR \ge 6$.



We want the distribution of $\chi^2$ for injected signals be separated as much as possible from the 
distribution of $\chi^2$ for the spurious events. The optimal number of bins is the one which corresponds to
the minimum overlap between those two distributions. One can see that for the case considered above the 
optimal number lies somewhere close to 20. We need a more rigorous way to define the optimal number of bins,
so that we need to quantify the overlap between the two distributions. Here we will apply the standard
detection technique \cite{Kay}. First we need to normalize the
distributions $P_1(\chi,N)$ (corresponds to the distribution of $\chi^2$ for the injected signals) and $P_2(\chi,N)$
(corresponds to the distribution of $\chi^2$ for the spurious events) so that
\bea
\int_{0}^{+\infty} P_1(\chi) d\chi = 1, \;\;\;\;\;\;\; \int_{0}^{+\infty}P_2(\chi) d\chi =1.
\ena
$P_2(\chi)$ defines the false alarm probability distribution function, so that we can fix the false alarm probability according to

\bea
\int_{0}^{\eta(N)}P_2(\chi,N)d\chi = \alpha.
\ena
By fixing the false alarm probability $\alpha$, we are essentially fixing the threshold, $\eta(N,\alpha)$, on $\chi^2$. Note that the threshold is a function of the number of bins and the false alarm probability. 
For real data, $\eta$ cannot be computed analytically, since $P_2$ depends on spurious events, or, 
rather on the similarity of spurious events to the chirp signal. Thus
the purpose of the playground data is to characterize the non-stationarities in the data.

{\it We will call the number of bins optimal if for a given $\alpha$ it maximizes the detection probability $P_d$}

\bea
\int_0^{\eta(N,\alpha)} P_1(\chi,N) d\chi = P_d.
\ena
 In other words,
\bea
N_{opt} = max_N \left( \int_0^{\eta(N,\alpha)} P_1(\chi,N) d\chi\right).
\label{opt}
\ena
Note, that we know $P_1(\chi,N)$ only for chirps plus Gaussian noise. The detector's noise, however, is not Gaussian over a long time scale, so that $P_1(\chi)$ is also, strictly speaking, unknown to us. This is why we
have used software injections. As a bonus we also derived a threshold on $\chi^2$, $\eta(N,\alpha)$, which should be used in the analysis of the full data set.

As one can see, this method can be applied to {\it any} signal based vetoing statistic. In Section~\ref{IV} we will
apply this method to determine the efficiency of other statistics.
As an example, we can apply Eq.~(\ref{opt}) to the simulation described
above and quantify the results presented in Fig.~\ref{chi21}.

\begin{table}[tbh]
\caption{\label{tab1}}
Optimization of $\chi^2$. Detection probability and threshold on $\chi^2$ for various number of bins.
False alarm probability in all cases was 1\%. 
\begin{ruledtabular}
\begin{tabular}{ccccccccc}
N bins    & 8    &  16    & 24     & 32    & 40     & 50     & 64     & 86   \\
$P_d$     & 58\% & 81.2\% & 85.4\% & 82\%  & 75.8\% & 68.3\% & 62.4\% & 47\% \\   
threshold & 1.59 & 8.985  & 20.8   & 34.54 & 47.46  & 65.97  & 95.11  & 143  \\
\end{tabular}
\end{ruledtabular}
\end{table} 
\begin{figure}[tbh]
\includegraphics[height=7cm, width=9cm,angle=0] {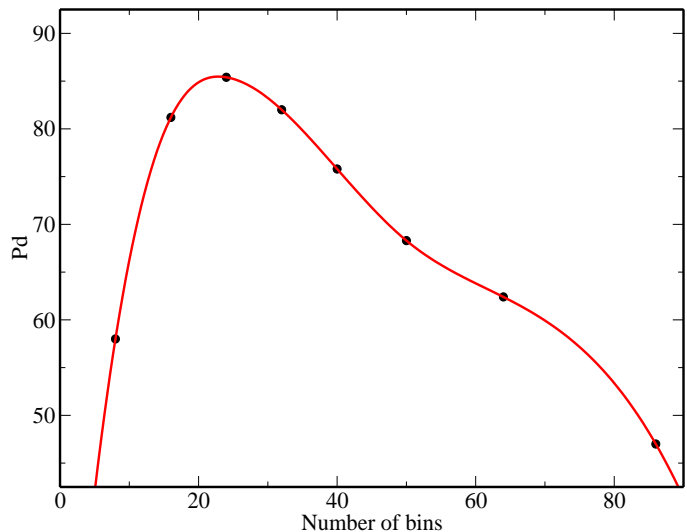}
\caption{Graphical representation of first two lines from the Table~\ref{tab1}. The solid line is a cubic spline interpolation.}
\label{splin}
\end{figure}

The results given in Table~\ref{tab1} (especially $P_d$) should be taken with caution. We have injected only 214 signals, and it 
might not be enough to make a definite statement. However it is a very good indication on what is the optimal number of bins. We have quite a large number (2550) of spurious events with $SNR \ge 6$, so that the statement 
about the threshold for a given false alarm probability is pretty solid. It should also be mentioned that we have truncated a tail of the $\chi^2$ distribution for spurious events by neglecting 5\% of all events with
largest $\chi^2$ (we continue 5\% truncation for false alarm distribution in the Section~\ref{IV} as well). 
We have also performed a cubic spline interpolation between these points (see Fig. \ref{splin}) to show that the optimal number of bins indeed lies somewhere close to 20.

At the end of this Section we would like to mention that an optimal parameter might not exist, or it
could be not the obvious one.

\section{Other signal based veto statistics}\label{IV}

  In this Section we will consider other signal based veto statistics. We start with a statistic that was inspired by the Kolmogorov-Smirnov ``goodness-of-fit'' test. We will show its statistical properties in the
 case of Gaussian noise. Then, we will consider some possible modifications of that statistic and another $\chi^2$-like statistic, which we will call $\hat{r}^2$. We show their performance using GEO\,600 S1 data.

\subsection{Kolmogorov-Smirnov based statistic}

\begin{figure}[tbh]
\includegraphics[height=6cm, width=6cm,angle=0] {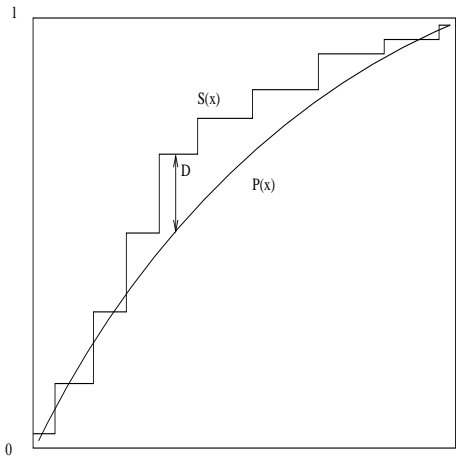}
\caption{Schematic representation of Kolmogorov-Smirnov test. Comparison between two cumulative 
distributions, $P(x)$ is a theoretical distribution and $S(x)$ is an observed one. Kolmogorov-Smirnov statistic is $D$.}\label{KSexampl}
\end{figure}

 The original Kolmogorov-Smirnov ``goodness-of-fit'' test 
 \cite{Kolmogorov, Smirnov, NR} compares two cumulative probability distributions, $S(x), P(x)$, 
(see Figure \ref{KSexampl}), and the test statistic is the maximum distance $D$ between curves $S(x)$ and $P(x)$.
 
Here we suggest a vetoing statistic which is somewhat similar to the Kolmogorov-Smirnov one, or better to say 
that the new statistic was inspired by the Kolmogorov-Smirnov test. We start by defining a few more quantities:

\bea
\psi_i &=& 2\sum_{f_k=F_0}^{F_i}\frac{\tilde{s}(f_k)\tilde{s}^*(f_k)}{S_n(f_k)}
\Delta f; 
\;\;\; i= 1,...,M, \;\;\; F_M= f_{lso}, \;\;\; \psi_M=1. \\
q_i &=& \sum_{f_k=F_0}^{F_i}\left( \frac{\tilde{x}(f_k)\tilde{s}^*(f_k)}{S_n(f_k)} + c.c.
\right) \Delta f; \;\;\;\; q_M=Q, 
\;\;\; y_k = \frac{\tilde{x}(f_k)\tilde{s}^*(f_k)}{S_n(f_k)}\Delta f,
\ena
where $M$ is defined by the frequency resolution.

The main idea is to compare two cumulative functions:
the cumulative signal power within the signal's frequency band and the cumulative SNR, which is essentially the correlation 
between the detector output and a template within the same frequency band.
Introduce the vetoing statistic according to 
\be
d= max_i\left|q_i - \psi_iQ\right|, \;\;\;\;\; i= 1,...,M-1
\label{d}
\en
and let us call it $d$-statistic.
However, we have found that, in practice, another statistic, $\hat{d}$: 
$$
\hat{d} = max_i\left| \frac{q_i}{Q} - \psi_i \right|,
$$
performs better. Nevertheless we start with $d$-statistic and postpone
consideration of $\hat{d}$ to the next subsection. The main question which we want to address is what is the probability of $d>D$ in the presence of a true chirp in Gaussian noise. Although we know that the detector's noise is not Gaussian, we can treat it as Gaussian noise plus non-stationarities (spurious transient events), and we try to discriminate those non-stationarities from the genuine gravitational wave signals. We refer the reader to Appendix~\ref{App1} for detailed calculations and we quote here only the final results.
If we introduce $Y_{i} = q_i - \psi_iQ$ (so that $d = max_i|Y_i|$), 
then the probability distribution function $P(Y_1,...,Y_{M-1})$ is the
multivariate Gaussian probability distribution function and

\bea
Pr(d>D) = 1 - \int_{-D}^{D} \frac{dY_1...dY_{M-1}}{(2\pi)^{(M-1)/2}\sqrt{det(C_{ij})}}
\exp\left( - \frac{{\bf YC}^{-1}{\bf Y}^{T}}{2}\right),
\ena
where the covariance matrix, ${\bf C}$, is defined in Eq.~(\ref{cov}).

\begin{figure}[tbh]
\includegraphics[height=8cm, width=10cm,angle=0] {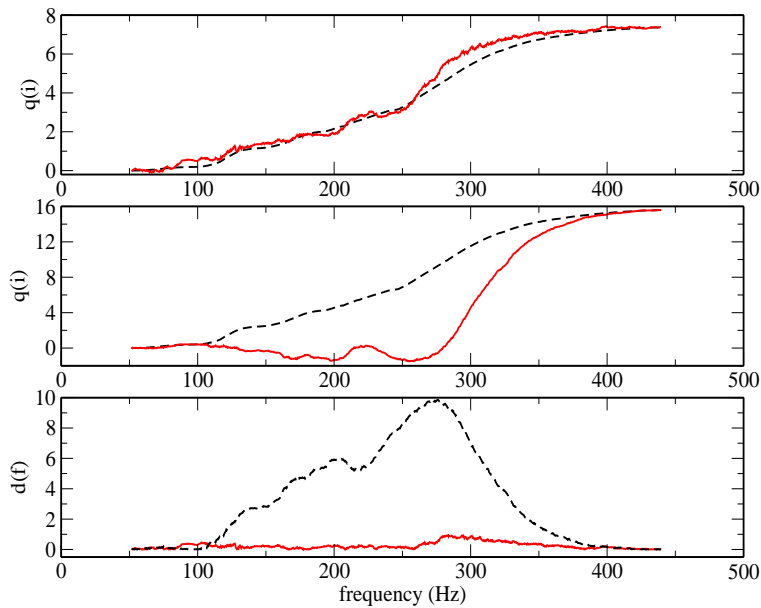}
\caption{Performance of $d$-test. Comparison of the cumulative SNR
versus expected (solid and dashed line correspondingly) for injected chirp
(the top graph) and for a spurious event (middle graph). The bottom plot shows distance $d$ as a function of frequency (the solid line is for injected chirp and the dashed line is for spurious event).}
\label{d_test}
\end{figure}

To show the performance of the $d$-test, we have computed $d$ for a glitch that produced SNR=16 at the output of the matched filtering and for the simulated chirp added to the data. The result is presented in the Fig.~\ref{d_test}.
The upper two panels show $q_i$: the top graph is plotted for a true chirp, and the middle graph is for a spurious event.
The dashed line corresponds to the expected cumulative SNR 
($\psi_i Q$) and the solid line is the actual accumulation ($q_i$).
The lower panel shows the distance ($|q_i - \psi_i Q|$) as a function of frequency. The solid line here corresponds to the
injected signal and the dashed line is for a spurious event.

As one can see, this test works in practice. However we have found that the
$\hat{d}$-statistic, defined above, performs better. One reason for this is that for the loud gravitational wave signals, we might have large $d$ due to slight mismatch in parameters caused by the coarseness of the template bank.

\subsection{Other vetoing statistics}

\begin{figure}[bht]
\includegraphics[height=9cm, width=10cm,angle=0] {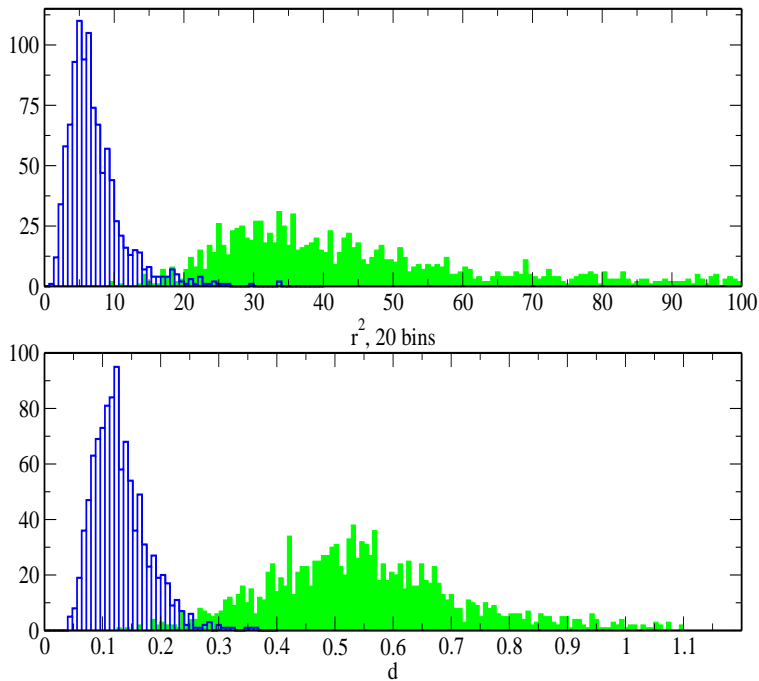}
\caption{Performance of $\hat{r}^2$ and $\hat{d}$ vetoing statistics 
are presented on the upper and lower plot correspondingly. The shaded  histogram corresponds to spurious events, and
the solid line histogram is distribution of vetoing statistic for injected signals. We have used one day's worth of S1 GEO\,600 data to conduct these simulations.}
\label{altrd1}
\end{figure}

We start with another $\chi^2$-like discriminator. The suggested statistic is

\bea
\hat{r}^2 = N \sum_{i=1}^{N} \frac{\left( \frac{Q_i}{Q} - p_i\right)^2}{p_i}.
\ena 
The interesting fact is that the TAMA group \cite{TAMAchi} is using a similar (related to the inverse of this quantity) statistic for the purpose of detection.  
In the following consideration we will omit the number of bins $N$ as it is just an overall scaling factor which does not
affect vetoing. One can see that $\chi^2$ introduced in Eq.~(\ref{chi2mod}) is related to the new statistic according to $\chi^2 = Q^2\hat{r}^2$.  
It is possible to derive the probability distribution function for $\Delta\hat{Q}_i = Q_i/Q -p_i$ for Gaussian noise following the same line as described in Appendix~\ref{App1}.
Unfortunately, the expression is quite messy, especially for the large number of bins $N$ and it is not very useful in practice. To check the performance of this statistic we have conducted simulations similar to the ones described in 
Section~\ref{III}. Namely, we have injected a chirp signal into a day's worth of S1 GEO\,600 data and plotted the two $\hat{r}^2$ distributions 
in the upper half of Fig.~\ref{altrd1}. 
The shaded histogram in the upper plot is a distribution of $\hat{r}^2$ for spurious events with $SNR\ge 9$ and the solid line curve is a distribution of $\hat{r}^2$ for injected chirp signals.
 We have chosen 20 bins to compute $\hat{r}^2$. Applying the scheme defined in the Section~\ref{III}, we find that the detection 
probability is 95.9\% and threshold is 16.47 for a false alarm probability of 1\%. Note that we did not use playground data for these simulations, so that our result might be biased by the choice of a particular data set.

Next, we will modify $d$-statistic according to

\be
\hat{d} = max_i \left| \frac{q_i}{Q} - \psi_i\right|.
\en
Define $\hat{Y}_i = q_i/Q -\psi$. We will skip the derivation of the probability distribution function 
$P(\hat{Y}_1,...,\hat{Y}_{M-1})$ in Gaussian noise.
As in the case of the $\hat{r}^2$ statistic, the probability 
could not be expressed in the nice close form, and, therefore,
is not useful in practical applications. 
The performance of $\hat{d}$ statistic is also shown in the Fig.~\ref{altrd1} (lower graph). To produce this picture 
we have used the same simulation as for $\hat{r}$. The detection probability for the $\hat{d}$-test is 94.3\% and the threshold 
is 0.21 for a false alarm probability of 1\%. 

Another possible modification of the $d$-statistic is choosing not the largest distance, but the percentile value,
in other words, the maximum distance after throwing away, say, 3\% of the largest distances. The percentile value could
be considered as a parameter for the $d$-statistic, and could be optimized for. 
To finish with $d$-like statistic, let us give a few other possibilities:

\bea
d^* &=& max_i \left| \frac{q_i/Q-\psi_i}{\sqrt{\psi_i(1-\psi_i)}}  \right|,\label{AD} \\
V = d_{+} + d_{-} &=& max_i \left( \frac{q_i}{Q}-\psi_i \right) + 
max_i \left( \psi_i - \frac{q_i}{Q} \right). \label{Kuiper} 
\ena
The first one, defined by Eq.~(\ref{AD}), is the analogue of Anderson-Darling \cite{AD} statistic and the second one, Eq.~(\ref{Kuiper}), is
 the analogue of Kuiper statistic \cite{Kuiper}.

\begin{figure}[tbh]
\includegraphics[height=9cm, width=13cm,angle=0] {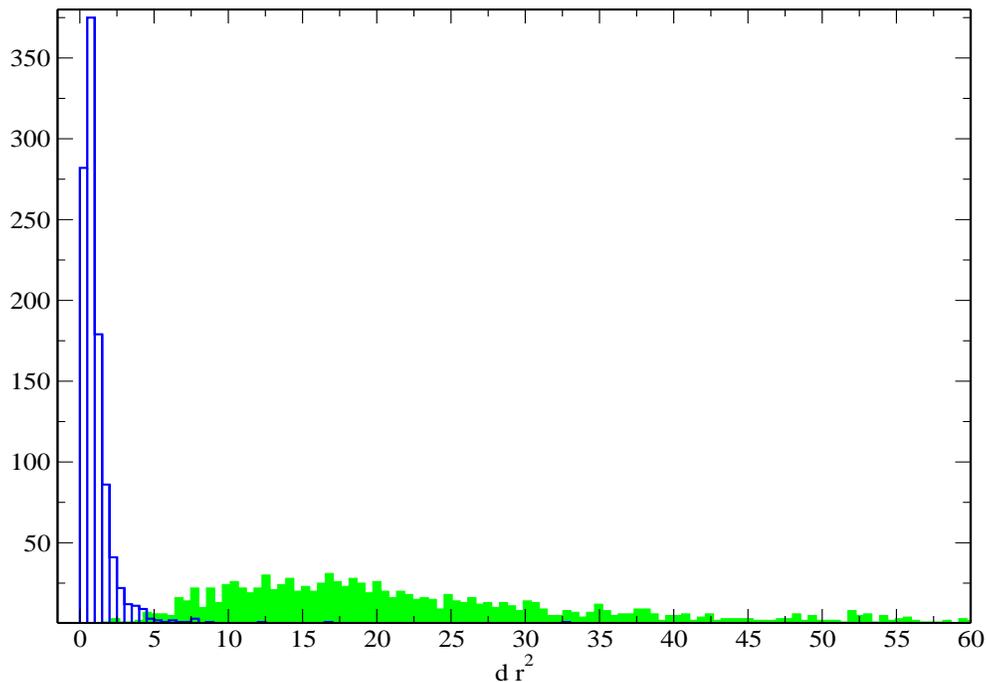}
\caption{Distribution of product statistic $(\hat{d}\times \hat{r}^2)$ for injected signals (solid line histogram) and
for spurious events (shaded histogram). We have used the same day-long
GEO\,600 data as for producing results presented in the Fig.~\ref{altrd1}. }
\label{prodct}
\end{figure}

The interesting fact is that the product of statistics $\hat{d}\times \hat{r}^2$ works even better
 than each of them separately and one can see this in the Figure~\ref{prodct}. The detection probability
in this case is 98.3\% and the threshold is 4.7 for a false alarm probability of 1\%. 

The reason that the product of two statistics works even better
than each of them separately could be because $\hat{r}^2$ and $\hat{d}$ might be 
better suited for different types of spurious events, and equally good for the true signals.
The statistics in the product supplement each other to veto larger number of spurious events.

We have tried to optimize $\hat{d}\times \hat{r}^2$ with respect to the number of bins, following the same line and conducting similar simulations as described in the Section~\ref{III}.
However, we have not found the obvious choice for the optimal number of bins. This is because the detection probability as a function of the number of bins for $\hat{d}\times\hat{r}^2$ fluctuates slightly about a constant value
for the number of bins between 18 and 40.


\section{Conclusion}\label{V}

In this paper we have considered several signal based vetoes. Those are various statistics based on our knowledge of the signal we search for, which help us in discriminating genuine gravitational wave
signal from spurious events of instrumental or environmental origin. 


We have outlined the method to optimize $\chi^2$-like statistic for the number of bins. This method is based on adding simulated signals to real data and studying the distribution of the vetoing statistic
for injected signals and spurious events. The optimal number of bins is the one which maximizes the detection probability for a fixed false alarm 
probablity. This method also automatically provides us with the vetoing threshold.

We have considered two other very promising signal based vetoes: $\hat{r}^2$ -- the $\chi^2$-like discriminator, and $\hat{d}$
-- the statistic which was inspired by the Kolmogorov-Smirnov ``goodness-of-fit'' test. Using again simulated injections into GEO\,600 S1 data we have shown that both those statistics could give a very high detection 
probability ($>94$\%) for a given false alarm probability (1\%). We have also pointed out that we can achieve even better performance if we take
the product of the two statistics as a new veto.

Finally, let us emphasize, that the results of the simulations presented here are data dependent, and the exact numbers for efficiency may vary for different detectors and/or for different data sets of the same detector. However, as it follows from the analytical evaluations and indicated from the conducted simulations, we should expect good performance for all signal based vetoes considered in this paper.

\acknowledgements
The research of S.~Babak was supported by PPARC grant PPA/G/O/2001/00485. 
S.~Babak would like to thank Bruce Allen and R. Balasubramanian for helpful and 
stimulating discussions. The authors would like to thank B.S. Sathyaprakash for helping to clarify the manuscript and for very useful 
suggestions and comments. Finally, the authors are also grateful to the GEO\,600 collaboration for making available the data taken by GEO\,600 detector during S1 run.

\appendix

\section{Derivation of statistical properties of $d$-test.}
\label{App1}

This Appendix is dedicated to deriving the probability that the $d$-statistic, introduced in (\ref{d}), is larger than a chosen value $D$.
The derivations presented here are conducted along the line similar to 
the one described in Appendix A of \cite{Ballen}.

We assume that the detector's noise $n(t_i)$ is Gaussian.
Introduce $Y_i = q_i -\psi_iQ$, then $d = max_i |Y_i|$. The main question we want to address is what
is the probability of $d>D$ in the presence of a true chirp:

\bea
Pr(d>D) &=& Pr(max_i \{|Y_i|\} >D) = 1 - Pr(max_i \{|Y_i|\} < D) = 
1-Pr(|Y_1|<D,...,|Y_{M-1}|<D) \nonumber\\ &=&  
1 -\int_{-D}^{D}...\int_{-D}^{D}P(Y_1,...,Y_{M-1})dY_1...dY_{M-1}.  
\ena
We need to find the probability distribution $P(Y_1,...,Y_{M-1})$ and we start with 
statistical properties of $y_k$:
$$
y_k = \left(\frac{\tilde{x}(f_k)\tilde{s}^*(f_k)}{S_n(f_k)} + 
c.c.\right) \Delta f
$$
We know that $y_k$ are $M$
independent Gaussian random variables. We can find their mean and variance,
\bea
E(y_k) &=& 2A\frac{\tilde{s}(f_k)\tilde{s}^*(f_k)}{S_n(f_k)} = A(\psi_k - \psi_{k-1}) \equiv A\phi_k,\\
var(y_k) &=&  \phi_k, \mbox{where we used notation  } \phi_k = 2\frac{\tilde{s}(f_k)\tilde{s}^*(f_k)}{S_n(f_k)}\Delta f.
\ena
Taking into account the fact that $y_k$ are independent and have normal distribution, 
${\mathcal N}(A\phi_k,\phi_k)$, we can write
\bea
P(y_1,...,y_M) = \prod_{i=1}^M \frac1{\sqrt{2\pi\phi_i}}\exp\left[ -\frac{(y_i-A\phi_i)^2}
{2\phi_i}\right].
\ena

We use the same trick as in \cite{Ballen}:

\bea
\int &dx_1...dx_{M-1}& \bar{P}(x_1,...,x_{M-1})F(x_1,...,x_{M-1}) = \\
\nonumber 
\int &dy_1...dy_{M}& P(y_1,...,y_M) F\left(y_1 - \psi_1\sum_{k=1}^{M}y_k,...,
\sum_{k=1}^{M-1}y_k - \psi_{M-1}\sum_{k=1}^My_k\right)
\ena
and choose $F(x_1,...,x_{M-1}) = \delta(x_1-Y_1)...\delta(x_{M-1}-Y_{M-1})$. This yields
\bea
P(Y_1,...,Y_{M-1}) = \int dy_1...dy_M P(y_1,...,y_M)\delta\left(y_1 - \psi_1\sum_{k=1}^My_k -Y_1\right) ...\nonumber\\
 \delta\left(\sum_{k=1}^{M-1}y_k - \psi_{M-1}\sum_{k=1}^My_k -Y_{M-1}\right).\label{int1}
\ena
Under the following change of variables of integration $(y_1,...,y_n) \to (z_1,...,z_{M-1},W)$
\bea
y_1 &=& z_1 + \phi_1W; \;\;\;\;\;  W= \sum_{k=1}^M y_k \nonumber\\
y_i &=& z_i - z_{i-1} + \phi_iW, \;\;\;\;\; i=2,...,M-1 \nonumber\\
y_M &=& -z_{M-1} + \phi_M W\nonumber\\
J &=& det \frac{\partial(y_1,...,y_M)}{\partial(z_1,...,z_{M-1},W)} = \sum_{k=1}^{M}\phi_k=1,\nonumber
\ena
the integral (\ref{int1}) takes the form

\bea
P(Y_1,...,Y_{M-1}) = \int dz_1...dz_{M-1}dW \left[ \prod_{i=1}^M \frac1{\sqrt{2\pi\phi_i}}
\exp\left[ -\frac{(y_i-A\phi_i)^2}{2\phi_i}\right] \right]\delta(z_1-Y_1)...\delta(z_{M-1}-Y_{M-1}).
\ena
The argument of the exponent can be expressed in term of new variables according to
\bea
\sum_{i=1}^M \frac{(y_i-A\phi_i)^2}{\phi_i} = \sum_{i=1}^{M}\frac{(z_i-z_{i-1})^2}{\phi_i} + (W-A)^2=
{\bf ZC}^{-1}{\bf Z}^T + (W-A)^2,
\ena
where in the expression above we used $z_0=z_M\equiv 0 $, {\bf Z} is a vector column  $(z_1,...,z_{M-1})$,
and ${\bf C}^{-1}$ is inverse of the covariance matrix, {\bf C},

\bea
{C^{-1}}_{ij} = \left( \frac1{\phi_i} + \frac1{\phi_{i+1}}\right)\delta_{ij} - \frac1{\phi_j}\delta_{i+1 j}
-\frac1{\phi_{i}}\delta_{i j+1}.\label{cov}
\ena 
Note that 
$$
det(C_{ij}) =  \frac1{det({C^{-1}}_{ij})} = \frac{\prod_{i=1}^M \phi_i}{\sum_{i=1}^{M}\phi_i} = 
\prod_{i=1}^M \phi_i.
$$
Taking all above into account and performing integration over $W$ we arrive at the required probability 
distribution function

\bea
P(Y_1,...,Y_{M-1}) = \frac1{(2\pi)^{(M-1)/2}\sqrt{det(C_{ij})}} \exp\left( 
- \frac{{\bf YC}^{-1}{\bf Y}^{T}}{2}\right),
\ena
which is the multivariate Gaussian probability distribution function. The final result can be written as

\bea
Pr(d>D) = 1 - \int_{-D}^{D} \frac{dY_1...dY_{M-1}}{(2\pi)^{(M-1)/2}\sqrt{det(C_{ij})}}
\exp\left( - \frac{{\bf YC}^{-1}{\bf Y}^{T}}{2}\right)
\ena
One also can compute mean and variance for each $Y_i$:
\bea
E(Y_i) &=& 0, \\
cov_{i\le j}(Y_iY_j) &=& \psi_i(1-\psi_j). 
\ena
We would like to emphasize, that like in the case or $\chi^2$ discriminator, $Y_{i}$ and, correspondingly $d$,
do not depend on the signal amplitude $A$.


\begin{thebibliography}{99}


\bibitem{LIGO} D. Sigg, {\it Class. Quantum Grav.} {\bf 21}, S409 (2004)
\bibitem{GEO} B. Wilke {\it et al}, {\it Class. Quantum Grav.} {\bf 21}, S417 (2004)
\bibitem{TAMARev} R. Takahashi {\it et al}, {\it Class. Quantum Grav.} {\bf 21}, S697 (2004)
\bibitem{VIRGO} F. Acernese {\it et al}, {\it Class. Quantum Grav.} {\bf21}, S385 (2004)

\bibitem{S1_1} The LIGO Scientific Collaboration, Abbott B., et al,
{\it Nucl. Instrum. Methods}, {\bf A517}, 154 (2004)
\bibitem{S1_2} The LIGO Scientific Collaboration, Abbott B., et al,
{\it Phys. Rev.} {\bf D69}, 122001 (2004)

\bibitem{TAMA} Takahashi H., Tagoshi H., the TAMA Collaboration,
{\it Class. Quant. Grav.}, {\bf 20}, S741 (2003)
\bibitem{TAMAchi} Takahashi H., Tagoshi H., et al, 
{\it Phys. Rev.} {\bf D70}, 042003 (2004) 

\bibitem{Ballen} Allen B., {\it Report gr-qc/0405045}.

\bibitem{PPN} Blanchet L., Faye G., Iyer B., and Joguet B.,
{\it Phys. Rev.} {\bf D65}, 061501(R), see also 064005 (2002)
\bibitem{EOB} Buonanno A., and Damour T., 
{\it Phys Rev.} {\bf D62}, 064015 (2000)
\bibitem{EOBspin} Damour T., {\it Phys. Rev.} {\bf D64},
124013 (2001)
\bibitem{DIS} Damour T., Iyer B and Sathyaprakash B.,
{\it Phys. Rev.} {\bf D63}, 044023 (2001)

\bibitem{Baggio} Baggio L., et al, {\it Phys. Rev.} {\bf D61},
102001 (2000)
\bibitem{Shawhan} Shawhan P. and Ochsner E., {\it Class. Quantum Grav.}
{\bf 21}, S1757 (2004)

\bibitem{Kolmogorov} Kolmogorov A., {\it Giorn. Inst. Ital. Attuari}
{\bf 4}, 83 (1933)

\bibitem{Owen} Owen B.J., {\it Phys. Rev.} {\bf D53}, 6749 (1996)

\bibitem{Cutler} Cutler C. and Flanagan E., {\it Phys. Rev.}
{\bf D49}, 2658 (1994)
\bibitem{Bala} Balasubramanian R., et al {\it Phys. Rev.}
{\bf D53}, 3033 (1996)

\bibitem{40m} Allen B., et al, {\it Phys. Rev. Lett.} {\bf 83},
1498 (1999)

\bibitem{Playground} LIGO 1 Collaboration, {\it Technical document
T030256-00}
\bibitem{Kay} Kay S.M., {\it Fundamentals of statistical signal processing. Detection theory},
Prentice Hall, (1993) 

\bibitem{Smirnov} Smirnov N.V., {\it Bull. Math. Univ. Moscow}, (1939)
\bibitem{NR} Press W.H., et al, {\it Numerical recipes in C}, Cambridge Univ. Press, (1992)
\bibitem{AD} Anderson T.W. and Darling D.A., {\it Anal. of Math. Stat.} {\bf 23}, 193 (1952)
\bibitem{Kuiper} Kuiper N.H., {\it Proceedings of Koninklijke Nederlandse Akademic van Wetenschappen}, ser A, {\bf 63}, 38 (1962) 

\end{thebibliography}
\end{document}